\documentclass{osa-article}

\journal{osac}

\articletype{Research Article}

\usepackage{graphicx}

\begin{document}

\title{Multimode Fiber Coupled Superconducting Nanowire Single Photon Detectors with High Detection Efficiency and Time Resolution}

\author{Jin Chang,\authormark{1,*} Iman Esmaeil Zadeh,\authormark{1} Johannes W. N. Los,\authormark{2} Julien Zichi,\authormark{2} Andreas Fognini,\authormark{2} Monique Gevers,\authormark{2} Sander Dorenbos,\authormark{2} Silvania F. Pereira,\authormark{1} Paul Urbach,\authormark{1} And Val Zwiller\authormark{3}}

\address{\authormark{1}Optics Research Group, ImPhys Department, Faculty of Applied Sciences, Delft University of Technology, Lorentzweg 1, 2628 CJ Delft, The Netherlands\\
\authormark{2}Single Quantum B.V., 2628 CH Delft, The Netherlands\\
\authormark{3}Department of Applied Physics, Royal Institute of Technology (KTH), SE-106 91 Stockholm, Sweden}

\email{\authormark{*}j.chang-1@tudelft.nl} 



\begin{abstract}
In the past decade superconducting nanowire single photon detectors (SNSPDs) have gradually become an indispensable part of any demanding quantum optics experiment. Until now, most SNSPDs are coupled to single-mode fibers. SNSPDs coupled to multimode fibers have shown promising efficiencies but are yet to achieve high time resolution. For a number of applications ranging from quantum nano-photonics to bio-optics, high efficiency and high time-resolution are desired at the same time. In this paper, we demonstrate the role of polarization on the efficiency of multi-mode fiber coupled detectors, and show how it can be addressed. We fabricated high performance 20, 25 and 50 $\mu$m diameter detectors targeted for visible, near infrared, and telecom wavelengths. A custom-built setup was used to simulate realistic experiments with randomized modes in the fiber. We simultaneously achieved system efficiency >80\% and time resolution <20 ps and made large detectors that offer outstanding performances.   
\end{abstract}

\section{Introduction}
Generating and detecting light at the single photon level has enabled a wide range of scientific breakthroughs in several fields, such as quantum optics, bio-imaging and astronomy. High performance single-photon sources have been realized in several platforms such as nonlinear crystals \cite{nonlinear}, color centers \cite{diamond}, atoms \cite{atoms}, molecules \cite{molecular}, and quantum dots (QDs) \cite{QDs}. Collecting light from most of these single-photon emitters, however, is a challenge. For example, QDs have emerged as excellent sources of single photons with outstanding single-photon purity \cite{Lucas} and promising candidates for high throughput generation of entangled photons \cite{entanglement1,entanglement2}. Currently, most high performance QDs are realized on III-V semiconductor platforms. Due to the nature of their emission and also high refractive index of these materials, extracting photons and coupling them to single-mode fibers has been a major challenge.  Many groups have explored processing of III-V quantum dots to enhance the coupling to optical fibers \cite{QD1,QD2,QD4}. However, coupling photons from QDs to single mode fiber, which has a low numerical aperture (NA) and a small core diameter, imposes demanding constraints on the laboratory setup. Multimode fibers, on the other hand, offer larger core diameters as well as higher NA, and provide several optical modes which significantly relaxes the task of optical coupling. In the field of tissue imaging, photons rapidly get scattered. After a short distance in the tissue, in the order of 1 mm, the transmitted ballistic light is attenuated by about 10 orders of magnitude \cite{tissue}. Therefore most collected photons are scattered and diffused ones. These photons cannot be easily collected and supported by single mode fibers thus large core, high NA and many modes are necessary. Similarly, in remote laser ranging applications, for example SNSPD-based Lidar system, multimode fibers are preferably chosen since they offer larger active area and easier coupling to telescope compared with single mode fiber. However, the SNSPDs used in the Lidar sysytem are smaller than the core size of multimode fiber \cite{lidar}. Thus large size SNSPDs coupled to multimode fiber with simultaneously high efficiency and high time resolution will benefit remote laser ranging applications.\par
Once light is coupled to fibers, SNSPDs are outstanding single-photon detectors because of their combined performances of high detection efficiency, high time resolution and low dark count rate\cite{Iman1}. Attempts have been made to couple SNSPDs to 50 $\mu$m and 100 $\mu$m, and using arrays of SNSPDs to 300 $\mu$m multimode fibers \cite{2014-visible,2015-50um,2017-100um,300um}. However, the achieved time resolution of these detectors have been limited to 76-105 ps.  Furthermore, it is not clear whether the mode-profile and also polarization of light was sufficiently random. \par    
In this paper, we designed and fabricated SNSPDs for several wavelength bands spanning from visible to telecom. To investigate the role of polarization on the system detection efficiency, we carefully characterized detectors using both standard SM fibers with polarization control and MM fibers with randomized modes and polarization. We also studied the influence of fiber dispersion on the instrument response function (IRF) of the system.

\section{SNSPDs fabrication and measurement setup}

\begin{figure}[hthp]
    \centering
    \includegraphics[width=9cm]{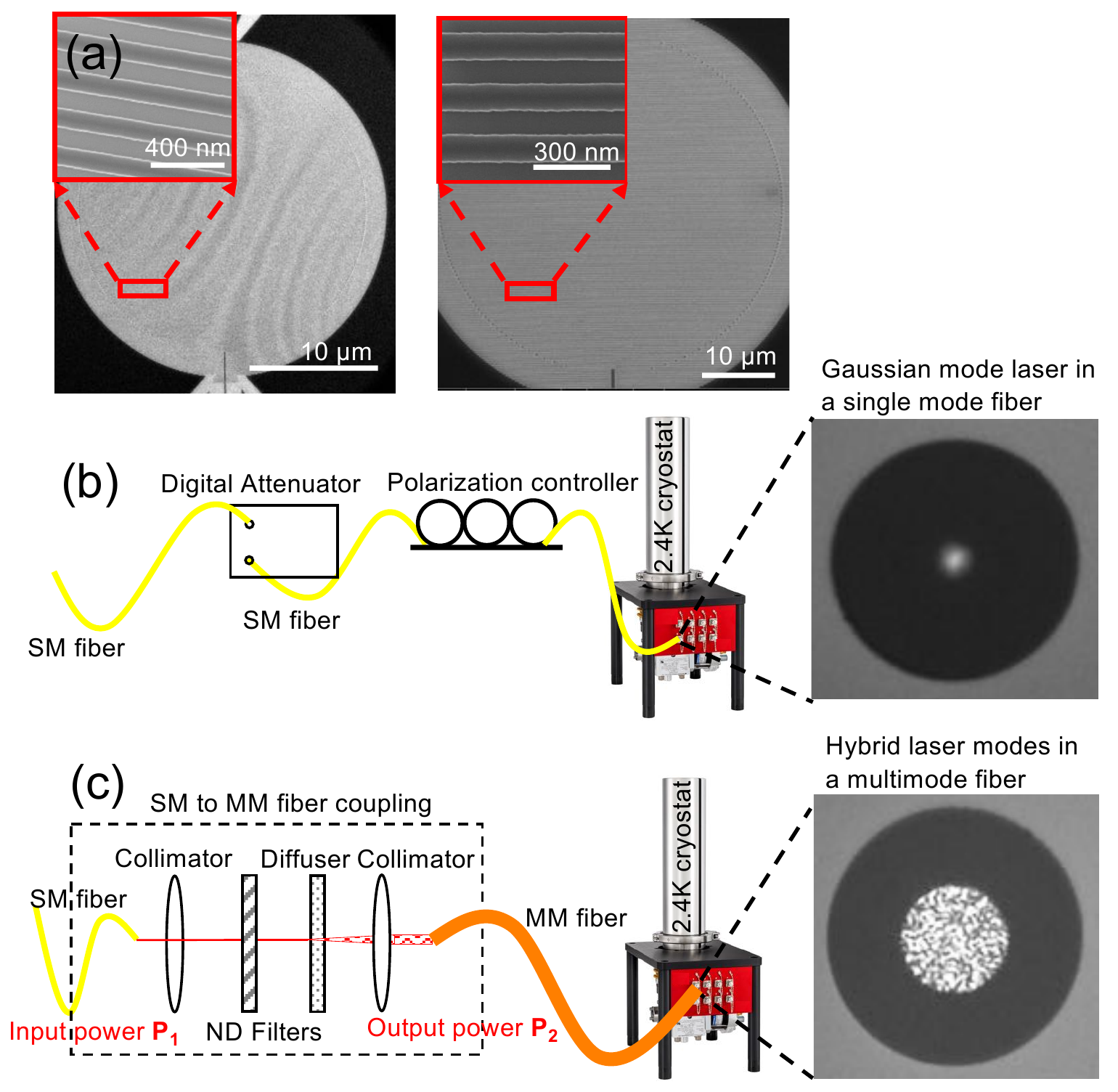}
    \caption{(a) Scanning electron microscopy images of a 20 $\mu$m diameter SNSPD (left) and 50 $\mu$m diameter SNSPD (right). System detection efficiency measurement with two different setups: (b) SM fiber setup and (c) MM fiber setup.}
    \label{fig:setup-SEM}
\end{figure}

Similar to [9], we fabricated SNSPDs for the wavelength range of 500-1550 nm out of sputtered 9-11 nm thick films of NbTiN. Figure \ref{fig:setup-SEM}(a) shows a scanning electron microscopy image of two fabricated SNSPDS, where left and right images show SNSPDs with 20 and 50 $\mu$m in diameter, respectively. To achieve saturation of internal efficiency, we fabricated the detectors for visible and near infrared using 10-11 nm films and the telecom SNSPDs out of 9 nm films. The meander width for visible detectors was 100 nm and for near infrared and telecom it was fixed to 70 nm. The filling factor in all cases was 0.5. 
As shown in figure \ref{fig:setup-SEM}(b) and (c), two optical setups were built to measure the SDE of SNSPDs with SM and MM fibers. In figure \ref{fig:setup-SEM}(b), light from a laser diode is guided by SM fiber to a digital attenuator, polarization controller (stress-induced birefringence produced by wrapping the fiber around three spools to create independent wave plates), and then coupled to another SM fiber which brings the light into the 2.4 K croystat. The right of figure \ref{fig:setup-SEM}(b)  shows the Gaussian mode of light inside the SM fiber. In order to evaluate the efficiency of multi-mode coupled devices for practical applications, as shown in figure \ref{fig:setup-SEM}(c), light from a laser diode is coupled to an optical U-shape bench containing neutral density (ND) filters for attenuating the light power and two diffusers with different grades for generating hybrid light modes. Another MM fiber is used at the other end of the U-bench to couple light into the 2.4 K cryostat. Inside the cryostat, the same type of MM fiber is used to couple light to SNSPDs. As a comparison, the picture on the right side of figure\ref{fig:setup-SEM}(c) represents an image of the core of MM fiber at the output of the U-bench indicating hybrid modes in the 50 $\mu$m multimode fiber.


\section{Simulation and system detection efficiency measurement}
\begin{figure}[hthp]
    \centering
    \includegraphics[width=13cm]{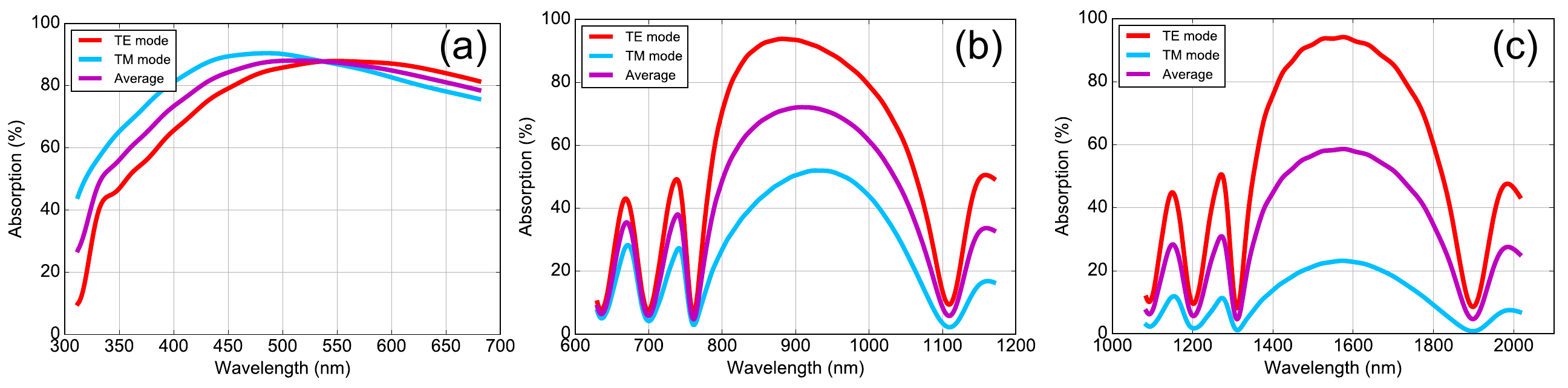}
    \caption{Simulated reflectivity of (a) Aluminum/$SiO_2$ cavity for visible wavelength (b) DBR for 900 nm and (c) DBR for 1550 nm wavelength.}
    \label{fig:simulation}
\end{figure}

Figure \ref{fig:simulation} shows the FDTD simulation of three different optical cavities for enhancing the photon absorption in the meander. Since the absorption is dependent on the  polarization of the vertically incident light \cite{polarization}, we simulated the maximum absorption (TE) and minimum absorption (TM) for Gaussian mode light. The average of TE and TM absorption, unpolarized (UP), is used to present the case where light is in hybrid mode. In figure \ref{fig:simulation}(a), an aluminum mirror with a thin layer of $SiO_2$  (about 70 nm) is employed to enhance the absorption for a meander with 100 nm line-width and 200 nm pitch. One benefit of this cavity is that the peak absorption wavelength can be simply shifted by tuning the thickness of the $SiO_2$ layer or adding extra $SiO_2$ on top. In figure \ref{fig:simulation}(a), the absorption wavelength center was set around 525 nm. At 516 nm, the absorption of TE mode reaches 89\% while for the TM mode it is 86\%. As a result, the averaged absorption of TE and TM remains at about 87.5\%. For 900 nm and 1550 nm, we simulated detectors on a Distributed Bragg Reflector (DBR) cavity for a better agreement with our experimental samples. In both cases, the DBR comprised 6.5 periods of $Nb_2O_5$ and $SiO_2$ bilayers. The thickness of $Nb_2O_5$/$SiO_2$ was 155/99 nm for 900 nm and 268/173 nm for 1550 nm, respectively. For both cases, we set the meander to be 70 nm line-width and 140 nm pitch, which was the same in our fabrication process. In figure \ref{fig:simulation}(b), the absorption of the TE mode approaches 92\% while TM mode is 52\% at 900 nm, and the averaged value is 72\%. At 1550 nm the simulated TE and TM efficiency, shown in figure \ref{fig:simulation}(c), are 92\% and 22\%, which yields and average of 57\%. This significant difference in absorption for TE and TM is due to stronger polarization dependence at telecom wavelength comparing to VIS and NIR. \par
\begin{figure}[hthp]
    \centering
    \includegraphics[width=13cm]{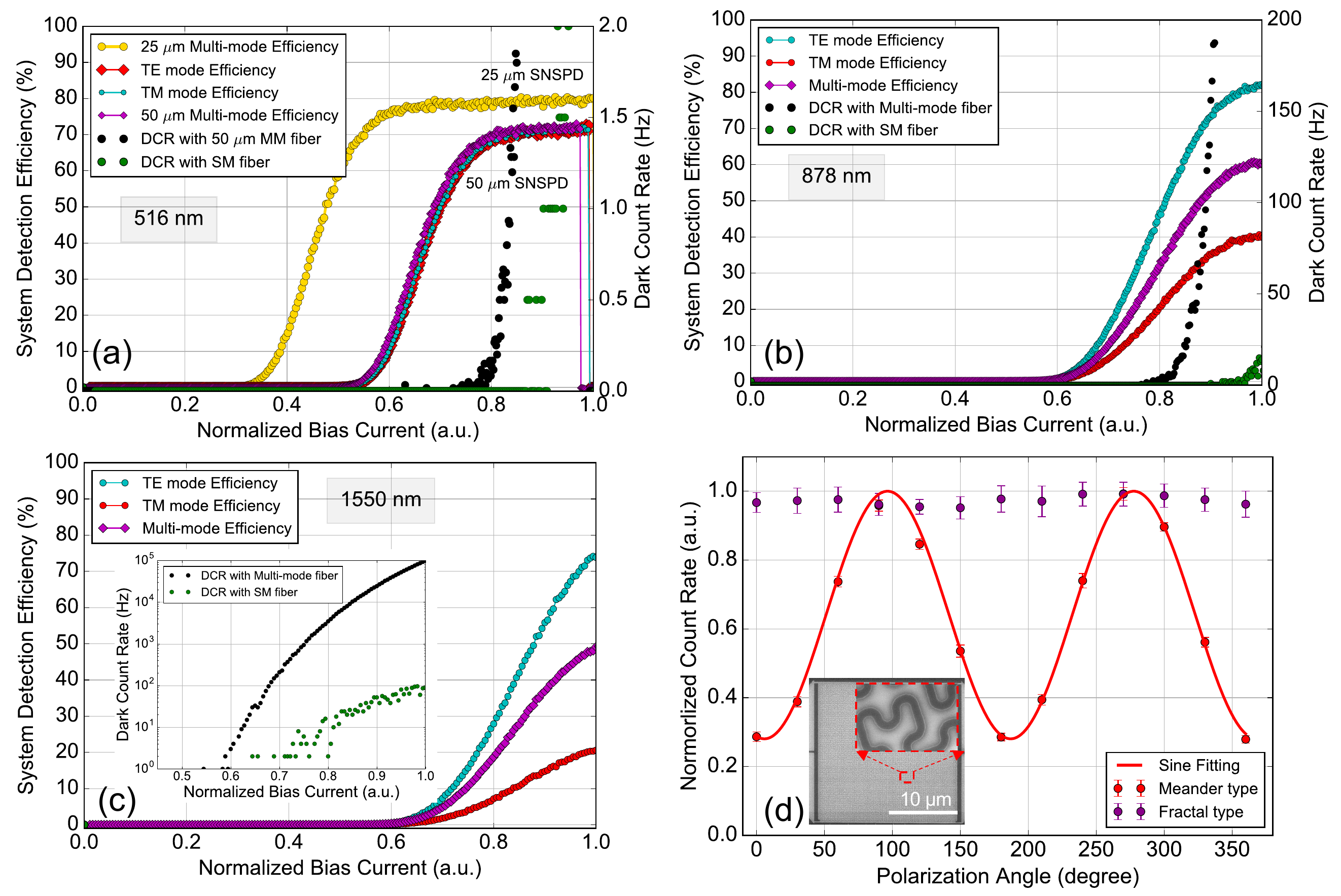}
    \caption{(a) SDE of  25/50 $\mu$m diameter SNSPD at 516 nm (b) SDE of a 20 $\mu$m diameter SNSPD at 878 nm (c) SDE of a 20 $\mu$m diameter SNSPD at 1550 nm and (d) Polarization dependence measurement of a meander detector (red dot) and a fractal detector (purple dot) at 1550 nm with 3-$\sigma$ error bar. Red curve shows sine fitting for the meander type detector.} 
    \label{fig:Eff-pulse}
\end{figure}

 
To characterize single mode SDE of our fabricated detectors, laser diodes with different wavelengths were used as photon source (continuous wave). The power of the laser was recorded by a calibrated power meter and was stabilized at 10 nW. After the power stabilization, an attenuation of 50 dB was added by the digital attenuator and then the laser was coupled to the SNSPD. For SDE measurement with MM fiber, we recorded the laser power before (P$_1$) and after (P$_2$) the U-bench with a calibrated power meter. By adding different ND filters in the U-bench, we controlled the ratio of P$_1$/P$_2$ close to 50 dB. Then we set the power of the input laser also at 10 nW for efficiency measurements. In both cases, the total input photon number can be back calculated from the input laser power. A commercial SNSPD driver was used to control the bias current and read the count rate. In all SDE measurements we subtracted dark counts from the total photon counts, and this was negligible for single mode measurements and multimode measurements at visible and the near infrared. However, telecom detectors coupled to MM fiber detect a significant amount of fiber coupled blackbody radiation. We also removed another 3.6\% to account for the end-facet reflection of the fiber where we measured the input power\cite{Iman1}. As shown in figure \ref{fig:Eff-pulse}(a), at 516 nm the SDE of the 50 $\mu$m detector is 70\% measured with both SM (red/cyan curve) and MM (purple curve) fiber. All detectors show well saturated internal efficiency. With SM fiber coupled to the SNSPD, there was negligible polarization dependence, which means the TE and TM modes are equally absorbed by the nanowiwe and therefore we also measured a similar efficiency with MM fiber coupled detectors. To avoid detector latching, we used a resistive bridge similar to \cite{Iman1}. We also fabricated 25 $\mu$m diameter SNSPDs (70 nm width/140 nm pitch) for visible wavelength range. Since this detector was significantly larger than the fiber, it offers more alignment tolerance and better absorption of cladding modes. An SDE about 80\% was achieved for this type of detectors coupled to a 20 $\mu$m fiber (yellow curve in figure \ref{fig:Eff-pulse}(a)). The dark count data shown in the figure is measured with both single mode and multimode fibers. When SNSPDs reach saturated detection efficiency, the dark count rate is below 0.2 Hz for both single mode and multimode fiber coupling at 516 nm, however, for 878 nm, figure \ref{fig:Eff-pulse}(b), and 1550 nm, figure \ref{fig:Eff-pulse}(c), dark count rates are much higher when SNSPDs are coupled to multimode fibers. As shown in the inset of figure \ref{fig:Eff-pulse}(c), when 1550 nm SNSPD is coupled to multimode fiber, the dark count rate is approximately 1000 times higher than the single mode fiber coupling. This is caused by black body radiation coupled to fiber modes and detected by efficient (compared to visible and NIR) telecom detector. 

At 878 nm as shown in \ref{fig:Eff-pulse}(b), TE and TM mode detection efficiency are significantly different. For optimized polarization (TE), the SDE approached 80\% while for TM polarization, SDE was only 40\%. The polarization dependence ratio was a factor of 2, which is in close agreement with our simulations. With our multi-mode efficiency measurement setup, we coupled hybrid modes and fully illuminate the SNSPD. We measured an SDE of 60\% for MM fiber coupled detector at 878nm which is in perfect agreement with an average of TE and TM efficiency measured with SM fiber. Similarly, at 1550 nm the TE efficiency was 75\% and TM was 20\%. The measured polarization dependence was larger, a factor of 3.75 and thus the efficiency measured with a multi-mode fiber was only 50\%. 

Polarization dependence can be improved by using index matching top dielectric layers \cite{luka1,luka2} at the cost of bandwidth. Fractal structures offer minimum polarization sensitivity over broad wavelength range \cite{fractal} so such devices are ideal for wideband multimode detection. We fabricated fractal structure SNSPD with 20 by 20 $\mu$m dimension as shown in the inset picture of figure \ref{fig:Eff-pulse}(d). Similar to \cite{sander-polarization}, we measured the polarization dependence for both meander and fractal type SNSPD by rotating a $\lambda$/2 wave plate over 360$^{\circ}$ (corresponding to 720$^{\circ}$ of the polarization direction). As shown in figure \ref{fig:Eff-pulse}(d), the red dot shows a strong polarization dependence of a meander type SNSPD (with 3-$\sigma$ error bar) and the measured data fits well with a sine function. On the other hand, the purple dots represent the measured polarization dependence of a fabricated fractal structure SNSPD and negligible (4\% difference between maximum and minimum) polarization dependence was observed.  

\section{Jitter measurement and analysis}
\begin{figure}[hthp]
    \centering
    \includegraphics[width=13cm]{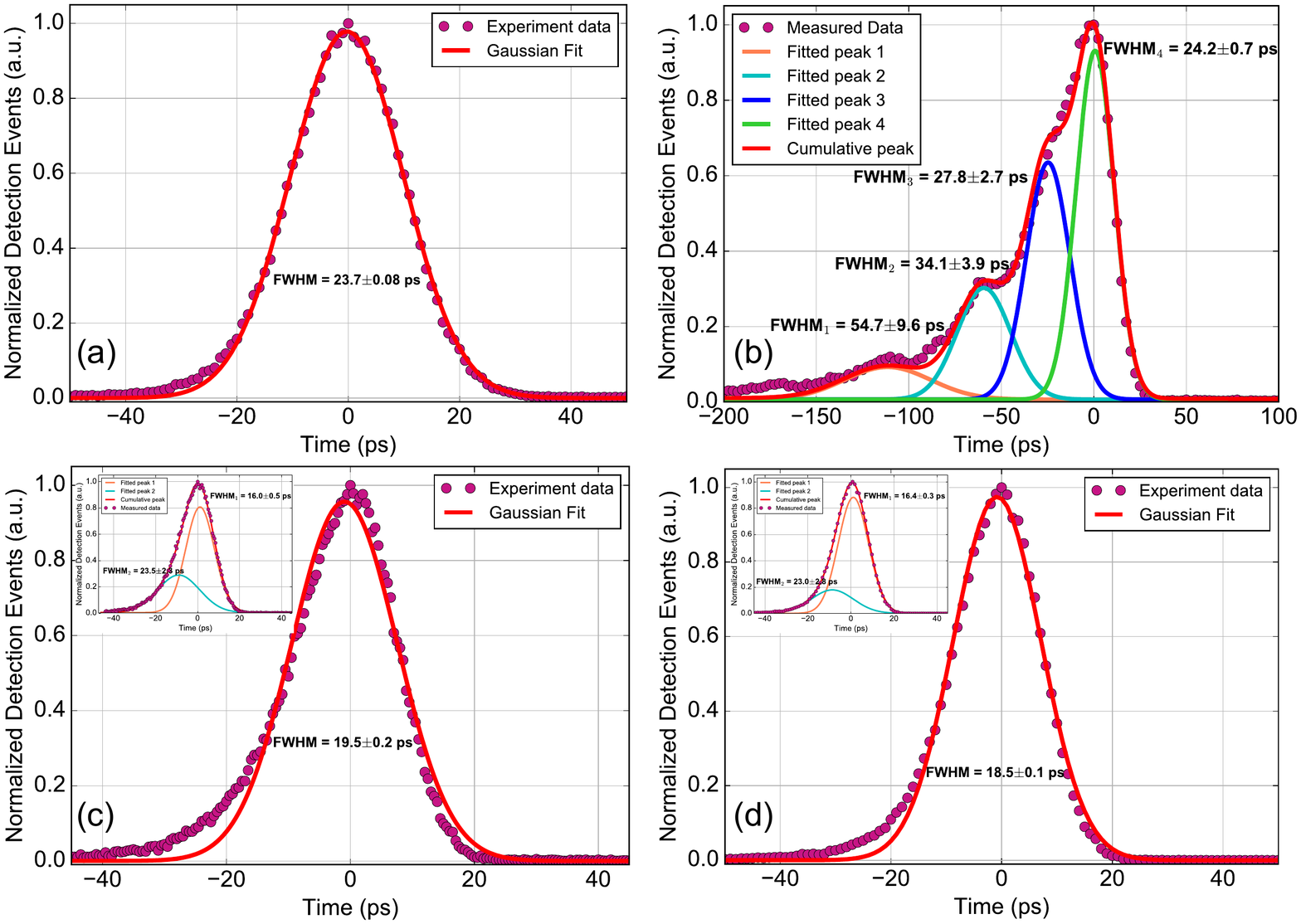}
    \caption{Jitter measurement of a 20$\mu$m diameter SNSPD with (a) graded-index MM fiber/RT amplifier (b) step-index MM fiber/RT amplifier (c) graded-index MM fiber/cryo amplifier and (d) SM fiber/cryo amplifier readout circuitry.}
    \label{fig:jitter}
\end{figure}

For SNSPDs, the instrument response function (IRF) usually has a Gaussian distribution and FWHM is commonly used as value of IRF of the system, also known as jitter. In this work, we used a 50 MHz picosecond pulsed laser (4.2 ps pulses) at 1064 nm and attenuated it to have much less than 1 photon per pulse ( on detector < 50-100 kHz count rate).  The electrical reference signal was provided by a fast photo diode. An oscilloscope with 4 GHz bandwidth and 40 GHz sampling rate was used to record the SNSPD signal pulse and to measure the jitter. We triggered on the rising edge of the SNSPD pulse as the start signal, and triggered the same way on the synchronized electric reference signal as stop. By building the distribution of time delay between start and stop, we acquired the IRF histogram and extracted its FWHM as jitter. In each jitter measurement we recorded over 100 thousands data points. We first measured the jitter of a 20 $\mu$m detector with a SM fiber and that yielded a value of 21.7 ps, then as shown in figure \ref{fig:jitter}(a), we measured the jitter of a 20 $\mu$m detector with graded index fiber (GIF) and room-temperature (RT) amplifier. The red curve represents the Gaussian fitting function and gives a jitter of 23.7$\pm$0.08 ps. In figure \ref{fig:jitter}(b), we used a step-index MM fiber inside the cryostat connecting to the same detector and also the same type of MM fiber as input. The jitter shows several peaks and by multi-peak fitting, we acquired the FWHM of each peak as shown in the figure. From left to right, they are 54.7$\pm$9.6, 34.1$\pm$3.9, 23.8$\pm$2.7 and 24.2$\pm$0.7 ps, respectively. The fitted peak 4 had the highest amplitude and 24.2$\pm$0.7 ps FWHM is close to the measured jitter with the SM fiber. However, since different modes travel at different speeds in the step-index fiber, known as dispersion, many other peaks appear. This makes graded index fiber the preferred choice for coupling to large diameter detectors. To reduce the contribution of the electrical noise mostly induced by the amplifier, we used a home-made cryogenic amplifier at the 40K stage. As shown in figure \ref{fig:jitter}(c), the jitter measured with GIF MM fiber and cryogenic amplifier shows Gaussian distribution and a FWHM of 19.5$\pm$0.2 ps. Similarly, in figure \ref{fig:jitter}(d) we show the jitter measured with SM fiber (smf28) and cryogenic amplifier and the FWHM was 18.5$\pm$0.1 ps. Since jitter has a noticeably asymmetric Gaussian shape in both figure \ref{fig:jitter}(c) and (d), in the inset of both figures two-peaks fitting is shown. According to a previous study \cite{jitter-theory}, in the deterministic regime where jitter is controlled by position-dependent detection threshold in straight parts of meanders, the main peak forms. And the lower side peak is caused in the probabilistic regime where the detector bends contribute. The achieved  jitter in this work, is the best reported time resolution for multimode fiber coupled SNSPDs so far.

\section{Conclusions}
In this paper, we designed, fabricated, and characterized multimode fiber coupled SNSPDs for visible, near infrared, and telecom wavelength. For visible wavelength, polarization dependence was negligible thus both single-mode and multi-mode coupled SNSPDs showed >80\% system detection efficiency. For near-infrared and telecom wavelengths, polarization dependence plays an important role and its ratio increases with wavelength. We reached 60\% and 50\% system detection efficiency with randomized mode illumination for 878 nm and 1550 nm, respectively. For jitter measurements, step-index multimode fibers introduce light dispersion thus the IRF showed multiple peaks. By using graded-index multimode fibers, this issue can be solved and together with  cryogenic amplifier readout circuitry, the jitter can be improved to 19.5$\pm$0.2 ps. Since our multimode fiber-coupled SNSPDs have high system detection efficiency and high timing resolution at the same time, they can improve the performance of existing experiments in quantum optics, life science and satellite based space communication in the future.

\section*{Acknowledgments}
J.C. acknowledges support from China Scholarships Council (CSC), No.201603170247. I.E.Z. acknowledges funding from the NWO LIFT-HTSM (680-91-202) grant.

\section*{Disclosure}
The authors declare that there are no conflicts of interest related to this article.



\bibliography{bibliography}
\end{document}